\documentclass[a4paper,fleqn]{cas-dc}

\usepackage[authoryear]{natbib}
\usepackage{lineno,hyperref,rotating,algorithm,algpseudocode}
\usepackage{bm,amssymb,multirow}
\usepackage{amsmath,amsfonts,color}
\usepackage{times}
\usepackage{graphicx}
\usepackage{url}
\usepackage{booktabs}
\usepackage{caption,subcaption}
\usepackage{array}
\usepackage{multicol}
\usepackage{multirow}
\usepackage{url,footnote}
\usepackage{stfloats}
\usepackage{makecell}
\usepackage{microtype,hyperref}

\def\tsc#1{\csdef{#1}{\textsc{\lowercase{#1}}\xspace}}
\tsc{WGM}
\tsc{QE}

\begin{document}
\let\WriteBookmarks\relax
\def\floatpagepagefraction{1}
\def\textpagefraction{.001}

\shorttitle{Physically Explainable CNN for SAR Image Classification}    


\title [mode = title]{Physically Explainable CNN for SAR Image Classification}  



%

\author[1]{Zhongling Huang}



\ead{huangzhongling@nwpu.edu.cn}



\affiliation[1]{organization={the BRain and Artificial INtelligence Lab (BRAIN LAB), School of Automation, Northwestern Polytechnical University},
            city={Xi'an},
            postcode={710072}, 
            country={China}}

\author[1]{Xiwen Yao}
\cormark[1]




\author[1]{Ying Liu}
\author[2]{Corneliu Octavian Dumitru}
\author[2,3]{Mihai Datcu}
\author[1]{Junwei Han}
\affiliation[2]{organization={Remote Sensing Technology Institute (IMF), German Aerospace Center (DLR)},
            city={Wessling},
            postcode={82234}, 
            country={Germany}}
\affiliation[3]{organization={University POLITEHNICA of Bucharest (UPB)},
            city={Bucharest},
            postcode={060042}, 
            country={Romania}}

\cortext[1]{Corresponding author}



\begin{abstract}
Integrating the special electromagnetic characteristics of Synthetic Aperture Radar (SAR) in deep neural networks is essential in order to enhance the explainability and physics awareness of deep learning. In this paper, we first propose a novel physically explainable convolutional neural network for SAR image classification, namely physics guided and injected learning (PGIL). It comprises three parts: (1) explainable models (XM) to provide prior physics knowledge, (2) physics guided network (PGN) to encode the knowledge into physics-aware features, and (3) physics injected network (PIN) to adaptively introduce the physics-aware features into classification pipeline for label prediction. A hybrid Image-Physics SAR dataset format is proposed for evaluation, with both Sentinel-1 and Gaofen-3 SAR data being experimented. The results show that the proposed PGIL substantially improve the classification performance in case of limited labeled data compared with the counterpart data-driven CNN and other pre-training methods. Additionally, the physics explanations are discussed to indicate the interpretability and the physical consistency preserved in the predictions. We deem the proposed method would promote the development of physically explainable deep learning in SAR image interpretation field.

\end{abstract}




\maketitle

\section{Introduction}

Synthetic Aperture Radar (SAR) can work in all-day all-weather conditions as an active microwave sensing technology. Different from the optical remote sensing images close to the visual understanding system of human eyes, SAR images reflect the electromagnetic characteristics of objects and terrain. In order to understand SAR images in a more comprehensive way, the artificial intelligence approaches should pay close attention on not only the visual information, but also the physical properties of SAR. 

SAR image classification is a basic task, aiming to assign the semantic label to each SAR image patch. Some conventional theory-driven approaches were explored to extract the hand-crafted features based on the expertise of SAR, e.g., the statistic model based \cite{leng2020fast,gao2017} and the physical model based methods \cite{leng2019}. These model based approaches have strong interpretability, yet the feature selection and classifier design are time-consuming and lack flexibility. As a comparison, the data-driven deep learning approaches can build an end-to-end system to learn the hierarchical features automatically and predict the semantic labels simultaneously without human intervention, superior to the pure model-based methods on SAR image classification tasks \cite{hzl,chen2016target}. 

Nevertheless, the current data-driven solutions for SAR image classification are still facing several challenges. The first is the contradiction between the data-hungry deep learning approaches and the expensive cost in manual annotation for SAR. At present, some pre-training related methods are popularized to tackle the issue, such as transfer learning and self-supervised learning. The transfer learning methods utilize the models pre-trained on other data domains (like natural images, optical remote sensing imagery, etc) via fine-tuning \cite{hzl}, domain adaptation \cite{huangwhat}, meta-learning \cite{Fu2022}, etc. The self-supervised learning usually takes the current domain data without annotations to optimize a designed contrastive or pretext task, obtaining the pre-training model for the downstream classification \cite{Wen2021,Ren2021}.

Despite the good performance of transfer learning and self-supervised learning on SAR image classification, the prediction of most deep models is short of physical explanation. In consideration of the special physical characteristics underlying SAR images, it is important to develop the hybrid approaches that blend the deep learning algorithms with physical models in SAR domain to keep the prediction consistent with physics and expert knowledge, which still remains a big challenge in the current researches. At first, such feature fusion methods that combine the CNN features with the hand-crafted description of SAR images were proposed \cite{Zhang2020,Wang2021,Sun2020}, but there still exist limitations on providing explanations and physical insights of SAR. More advanced hybrid approaches are required to embed the prior scientific knowledge from physical model into the deep neural networks \cite{de_B_zenac_2019}, particularly in SAR image classification domain, where the relevant studies are at the initial stage with a rising trend \cite{HUANG2020179}.

To meet the above challenges in SAR image classification, we propose a novel physically explainable CNN that blends the data-driven and model-driven approach to perform impressive generalization with limited labeled data and achieves physically explainable predictions. Our motivations are two-fold, as depicted in Fig. \ref{fig_intro}. Firstly, we intend to build a physics-inspired data-driven model for SAR image classification such as \cite{daw2020physics,PARK2019115883,8110836} in other research fields, which embeds the knowledge prior of physical model into the neural network. Secondly, inspired by self-supervised learning where the semantic feature embeddings are learned without supervision, we set a surrogate task based on the physical model to leverage the unlabeled SAR images and further support the main classification task. The proposed method, namely physics guided and injected learning (PGIL), is composed of three modules: 
 
\begin{figure}
     \centering
         \includegraphics[width=0.3\textwidth]{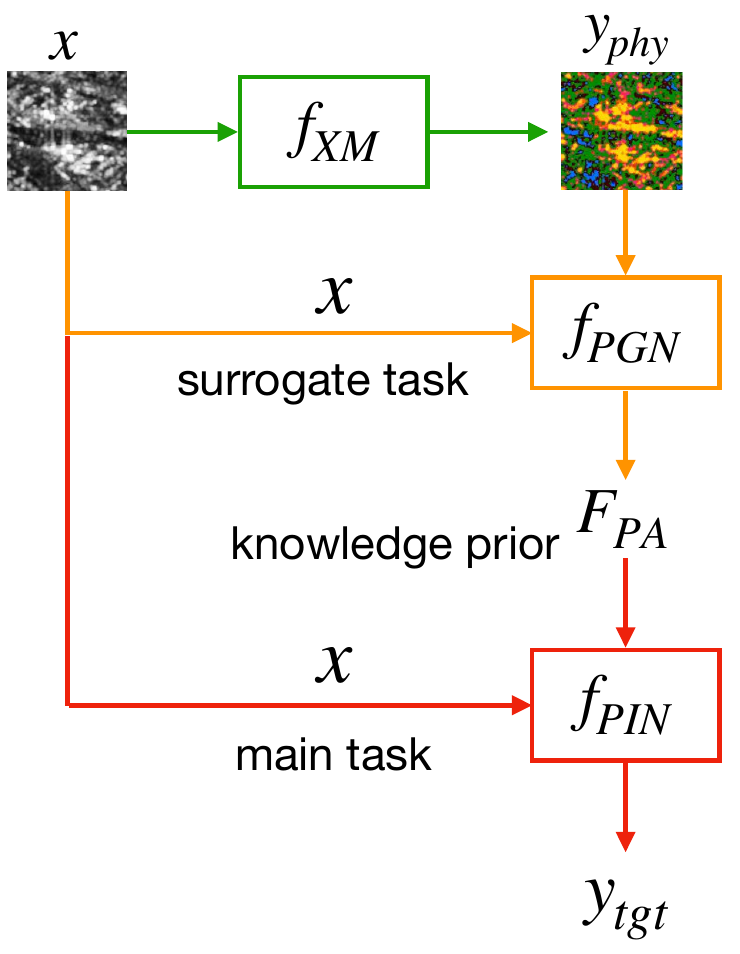}
     \caption{A surrogate task $f_{PGN}$ is built based on the physical information of SAR $y_{phy}$ derived from the explainable model $f_{XM}$. Thus, the prior knowledge of physical model is embedded as feature representation $F_{PA}$, which is successively injected into the main classification task via $f_{PIN}$ to learn the semantic label $y_{tgt}$.}
     \label{fig_intro}
\end{figure}

\begin{enumerate}
  \item \textbf{Explainable Model (XM):} We adopt the explainable model to generate the abstract physical representation for each SAR image patch in semantic level.

  \item \textbf{Physics Guided Network (PGN):} We propose a novel unsupervised learning neural network based on the designed surrogate task under the guidance of the abstract physical representation. The knowledge prior in the explainable model is converted as feature embeddings with PGN, aware of physical properties of SAR.

  \item \textbf{Physics Injected Network (PIN):} As for the main classification task, PIN is proposed to introduce the physics-aware features into the popular CNN pipeline. It captures more comprehensive representations and maintains the physical consistency of features, so as to prevent overfitting effectively with a few labeled samples available.
  
\end{enumerate}

For evaluation, a hybrid Image-Physics dataset format is proposed, equipped with both SAR image patches and the corresponding physical scattering mechanisms. Sufficient experiments are conducted mainly on the Sentinel-1 sea-ice classification dataset and also the Gaofen-3 SAR data to demonstrate the effectiveness of each module in the proposed method. The results show that the proposed PGIL exhibits remarkable generalization performance compared with the counterpart CNN architecture with supervised learning, transfer learning, and self-supervised learning in case of limited labeled data. More importantly, the physical explanations are discussed to demonstrate how the prior knowledge constrains the network from training and how the physics consistency is maintained in the predictions.

The contributions are summarized as follows:

\begin{enumerate}
    \item A novel physically explainable deep learning method is proposed for SAR image classification that deeply integrates the data-driven and theory-driven approaches. 
    
    \item By establishing a novel surrogate task based on explainable physical models, an unsupervised physics guided network is optimized to learn general features aware of prior knowledge.

    \item A hybrid Image-Physics dataset formation is proposed for evaluation, which combines the image and physics information of SAR in a concise way.
    
    \item We analyze the physics awareness of features, the good generalization on limited labeled data, and the explainability as well as the physics consistency of the predictions by sufficient experiments and discussions. 
\end{enumerate}

The rest of this paper is organized as follows. Section \ref{sec:bkg} reviews the background knowledge of physical models applied in this paper. Section \ref{sec:method} presents the physics guided and injected learning (PGIL) neural network for SAR image classification. The experiments and discussions are given in Section \ref{sec:exp}. Finally, Section \ref{sec:con} provides the conclusions.

\section{Background}
\label{sec:bkg}

In this section, we introduce the background of explainable theory-driven models for SAR applied in the following proposed method.

\begin{figure*}
     \centering
     \begin{subfigure}[b]{0.3\textwidth}
         \centering
         \includegraphics[width=\textwidth]{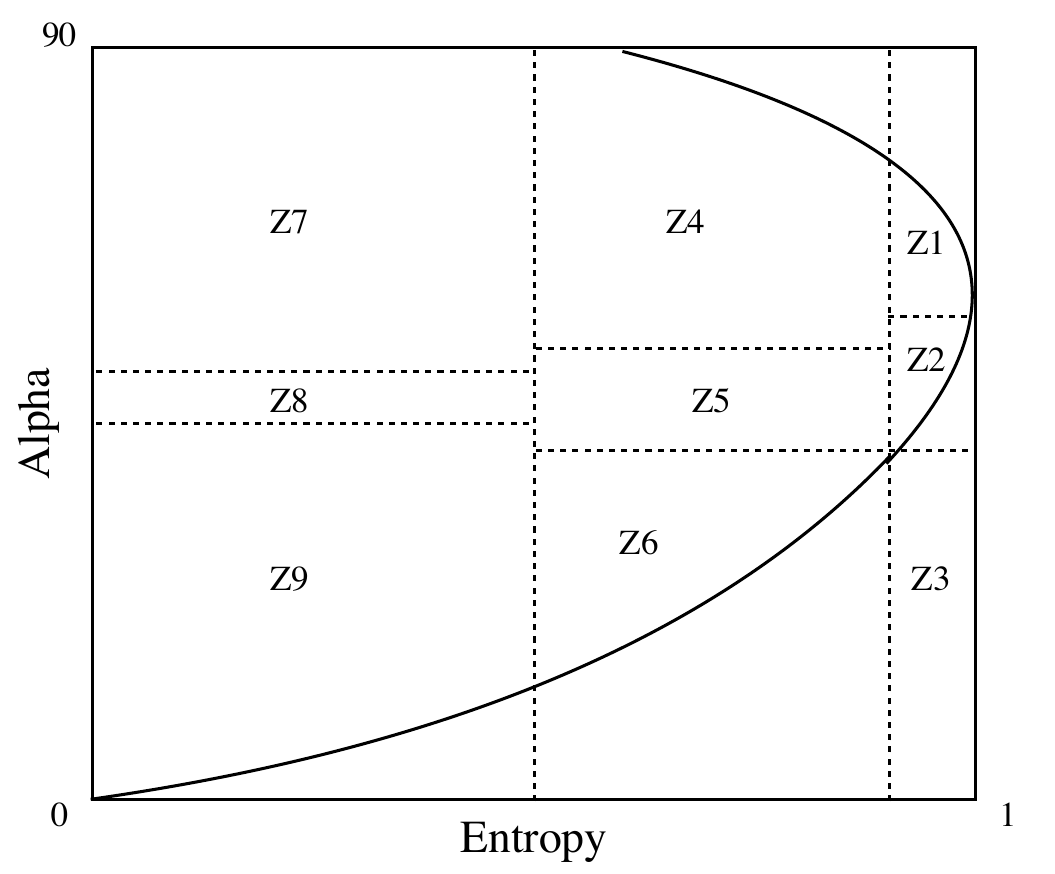}
         \caption{}
     \end{subfigure}
     ~
     \begin{subfigure}[b]{0.6\textwidth}
         \centering
         \includegraphics[width=\textwidth]{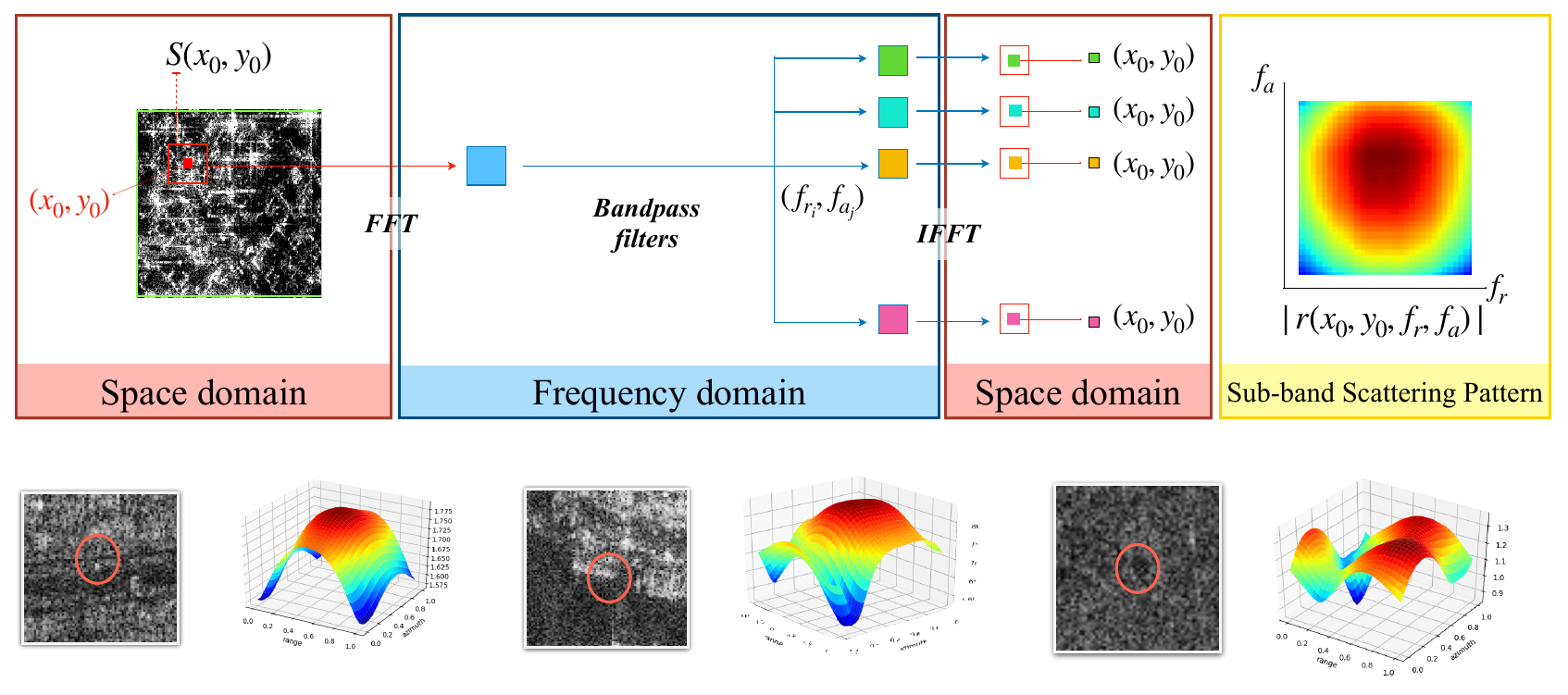}
         \caption{}
     \end{subfigure}
     \caption{Two different physical models of SAR. (a) shows the 2-dimensional H/$\alpha$ classification space \cite{Cloude1997} to demonstrate the scattering mechanisms for full-polarized SAR data. (b) indicates the time-frequency analysis model in HDEC-TFA method \cite{Huang2020} where the backscattering variations in different range and azimuth bandwidths of SAR targets are characterized.}
     \label{fig_bkg}
\end{figure*}

The first is the target decomposition model for PolSAR data to represent the target scattering by several basic scattering mechanisms. One of the well-known methods is the Cloude-Pottier decomposition for full-polarized SAR \cite{Cloude1997}, with the entropy $H$ and the angle $\alpha$ calculated from coherency matrix. An $H/\alpha$ plane is separated into nine zones to depict different scattering characteristics of full-pol SAR data, as shown in Fig. \ref{fig_bkg} (a). The scattering mechanism classification result can be obtained via the complex Wishart classifier proposed in \cite{lee1999}. Afterwards, the Cloude-Pottier decomposition model has been improved for dual-polarized SAR images \cite{Ji2015}. In this paper, we employ the Cloude-Pottier decomposition for both full- and HH/HV dual-pol SAR data \cite{Cloude1997,Ji2015}, and the scattering mechanism classification results are obtained by SNAP software.

The polarimetric decomposition is no longer available in single channel SAR image data. The second one we introduce in this paper is based on the time-frequency analysis model for single-polarized SAR data, as shown in Fig. \ref{fig_bkg}(b). The 2-dimensional short-time Fourier transform based time-frequency analysis on complex-valued high resolution SAR data characterizes the backscattering intensity variations of targets with different range and azimuth bandwidths, denoted as sub-band scattering pattern \cite{Huang2020}. Given a specific target with the position of $(x_0,y_0)$, and a segment $s$ centered in $(x_0,y_0)$. The sub-band scattering pattern of target $(x_0,y_0)$ is defined as
\begin{equation}
r(x_0,y_0,f_r,f_a)=abs(\mathrm{FFT}^{-1}\{\mathrm{FFT}(s) \cdot w(f_r, f_a)\}(x_0,y_0)),
\end{equation}
where $w(f_r, f_a)$ represents a series of bandpass filters centered on frequency pairs $\{(f_r, f_a)\}$ in both range and azimuth directions. The details can be found in \cite{Huang2020}. Fig. \ref{fig_bkg}(b) gives some examples of the extracted sub-band scattering pattern for different targets. A learning based HDEC-TFA method was proposed in \cite{Huang2020} to classify the scattering patterns.

\begin{figure}
     \centering
     \begin{subfigure}[b]{0.23\textwidth}
         \centering
         \includegraphics[width=\textwidth]{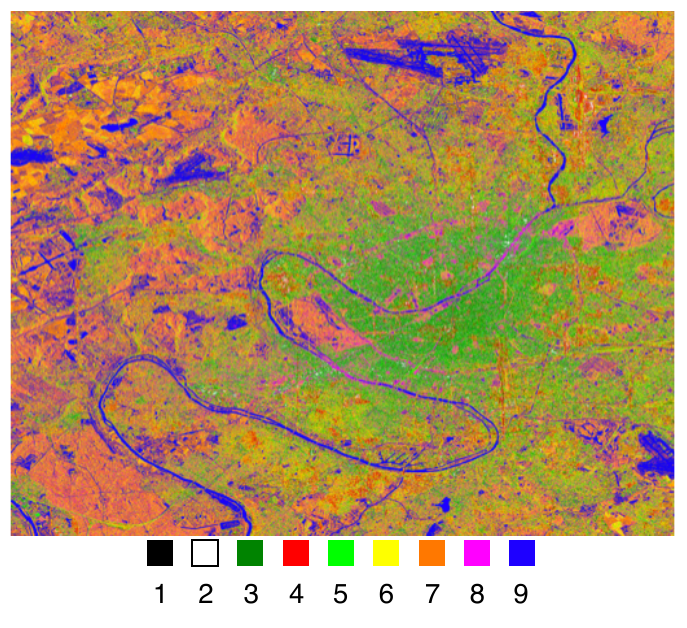}
         \caption{}
     \end{subfigure}
     ~
     \begin{subfigure}[b]{0.23\textwidth}
         \centering
         \includegraphics[width=\textwidth]{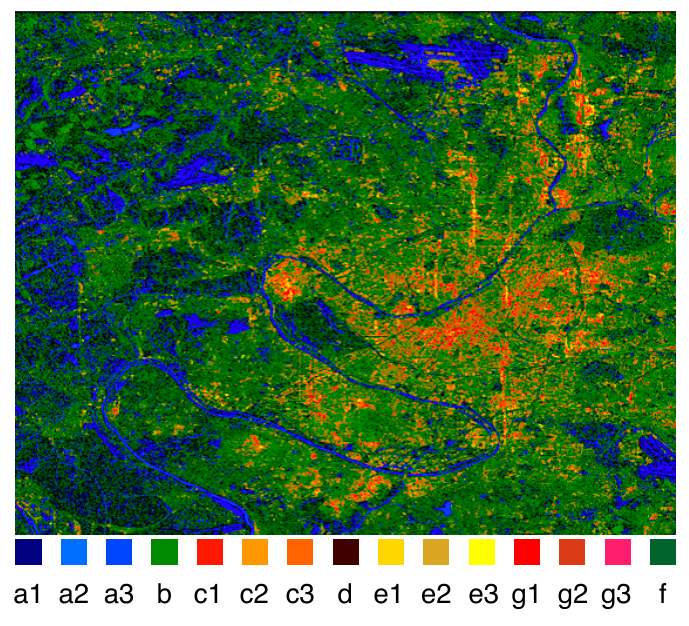}
         \caption{}
     \end{subfigure}
     \caption{Given a SAR image $x$, (a) shows the H/$\alpha$-Wishart classification result \cite{Cloude1997} on full-polarized SAR data, and (b) is the HDEC-TFA result \cite{Huang2020} on single channel (HH) SAR data.}
     \label{fig_Lxexp}
\end{figure}

The above two different physical models will be applied in our proposed method. Fig. \ref{fig_Lxexp} (a) shows the visualized result of H/$\alpha$-Wishart classification \cite{Cloude1997} on full-polarized SAR data, where the labels are consistent with Fig. \ref{fig_bkg} (a). Fig. \ref{fig_Lxexp} (b) is the HDEC-TFA result \cite{Huang2020} on single channel (HH) SAR image, where the given classes and colorization are in accordance with \cite{Huang2020}.

\section{Physics Guided and Injected Learning}
\label{sec:method}

\subsection{Overview}

In most patch-wise SAR image classification methods, the processed SAR amplitude images denoted as $x_I$ are considered other than the original complex product $x$ to predict the image label $y_{tgt}$. Thus, the intrinsic electromagnetic characteristics of SAR is not considered but desired. To this end, our proposed physics guided and injected learning (PGIL), as summarized in Fig. \ref{fig_intro}, leverages the visual friendly image data and the underlying prior knowledge in physical model. The basic motivation is to embed the physics knowledge into the neural network effectively.

\begin{figure*}[!t]
\centering
\includegraphics[width=13cm]{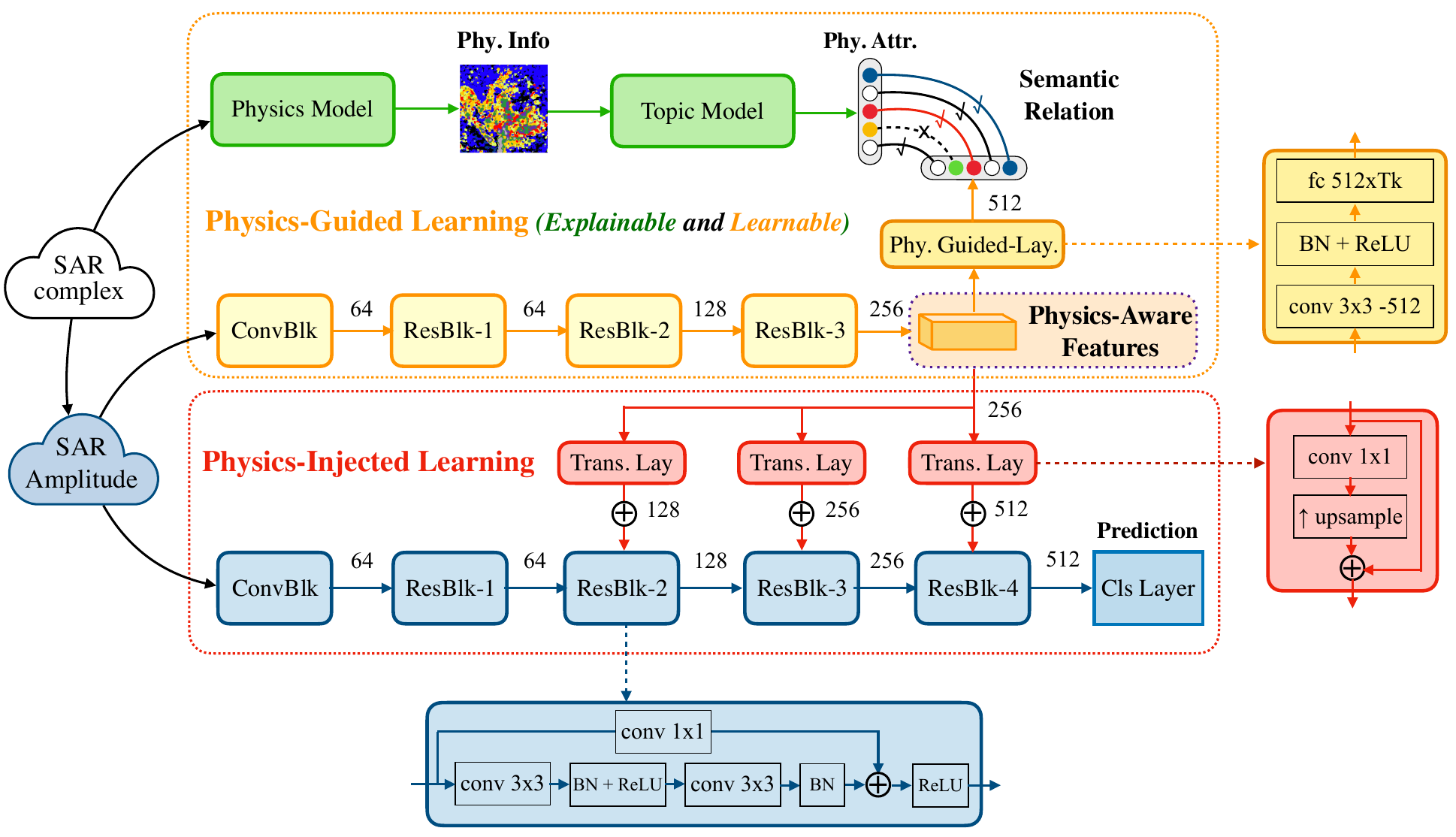}
\caption{The physics guided and injected learning (PGIL) for SAR image classification.}
\label{fig_method}
\end{figure*}

Three main modules are included in PGIL, that are explainable models (XM), physics guided learning network (PGN), and physics injected learning network (PIN). XM offers prior knowledge of physical model. PGN convert the prior knowledge into feature embedding, which is successively fused in PIN for label prediction. The overall framework is depicted in Fig. \ref{fig_method}.

XM acts on complex SAR image data $x$, where the explainable descriptor $y_{phy}$ is obtained to represent the physical scattering properties of SAR image:

\begin{equation}
\label{equ:pdm}
f_{XM}: x \to y_{phy}.
\end{equation}

$y_{phy}$ plays a major role in establishing the surrogate task of PGN for optimization, therefore referred to physics guided signal. PGN follows an unsupervised learning manner and outputs the feature embedding $F_{PA}$ aware of prior physical knowledge, namely physics-aware features. The mapping function $f_{PGN}$ is written as:

\begin{equation}
\label{equ:pgnn}
f_{PGN}: \{x, y_{phy}\} \to F_{PA}.
\end{equation}

Finally, PIN is proposed to complete the main classification task where the physics-aware feature $F_{PA}$ is injected, denoted as

\begin{equation}
\label{equ:pinn}
f_{PIN}: \{x, F_{PA}\} \to y_{tgt}.
\end{equation}

\subsection{Explainable Models}

As introduced in Section \ref{sec:bkg}, the physics-based H/$\alpha$-Wishart \cite{Cloude1997,Ji2015} and HDEC-TFA \cite{Huang2020} models serving as a part of XM, are adopted to obtain the scattering mechanism of target in SAR image $x$, denoted as $\mathcal{L}(x)$. The discrete physical scattering labels $\mathcal{L}(x)$ either depict different zones in H/$\alpha$ plane \cite{Cloude1997,Ji2015}, or refer to different targets with diverse scattering variation patterns (Fig. \ref{fig_bkg}(b)) \cite{Huang2020}.

Compared with SAR image label $y_{tgt}$, $\mathcal{L}(x)$ is too physics-specific to offer the semantic information. In view of optimizing PGN to obtain the feature embedding $F_{PA}$ aware of physics knowledge as well as being semantic distinctive, we additionally employ the Latent Dirichlet Allocation (LDA) on $\mathcal{L}(x)$ to output the topic mixture $y_{phy}$, denoted as $y_{phy} = \mathcal{G}(\mathcal{L}(x))$.

In LDA topic modeling, the \textit{document} formulated with bag-of-words representation is characterized by a distribution over latent topics, and each topic is represented by a distribution over \textit{words} in the \textit{vocabulary}. The \textit{corpus} is gathered from dataset to train the LDA model in an unsupervised pattern and the generative process is explainable. The details can be found in the related literature \cite{lda,lda4image2013}. We redefine the essential variables of LDA on the basis of SAR scattering characteristics as follows.

\begin{itemize}

  \item \textbf{word vector:} \qquad Randomly crop an area with the size of 8$\times$8 from $\mathcal{L}(x)$ as $\mathcal{L}_i$ and calculate the normalized histogram of $\mathcal{L}_i$ as the word vector.

  \item \textbf{vocabulary:} \qquad With randomly generated word vectors, apply the $k$-means algorithm \cite{kmeans} to obtain the vocabulary $V$ with $N_v$ cluster centers.

  \item \textbf{document:} \qquad For each SAR image patch, the set of scattering \textit{word vector} is gathered by tiling $\mathcal{L}(x)$ with a step-size of 4. The \textit{document} is given by the frequency of each word in \textit{vocabulary} $V$.

  \item \textbf{corpus} \qquad The corpus is collected for training the LDA model that formed as a matrix with the size $N_v \times N_d$, where $N_d$ is the number of \textit{document}.

\end{itemize}

Finally, $y_{phy}$ is obtained as the topic mixture (namely Bag of Topics, BoT) of $\mathcal{L}(x)$,
\begin{equation}
\label{equ:bot}
\mathcal{G}(\mathcal{L}(x)) = \{\varphi_1, \varphi_2, ..., \varphi_K\},
\end{equation}
where $\varphi_k$ denotes the score of the $k$th topic. Generally, the summation of $\varphi_k$ equals 1.

\subsection{Physics Guided Network}

The role of PGN lies in embedding the prior physics knowledge in a neural network, so as to extract the physics-aware features with semantic discrimination beneficial to classification. The optimization of PGN is motivated by the pretext task setting in self-supervised learning \cite{Misra_2020_CVPR}.

Tănase et al. \cite{radu} pointed that the topic semantics of scattering properties are close to human semantics used for basic land-cover types. Correspondingly, we propose to build a surrogate task under the following assumption: \textit{the SAR image features and the topic mixture of physical scattering labels should share common attributes in semantic level.} In other word, the physics descriptor $y_{phy}$ can be \textbf{partly} represented by high-level deep features extracted from SAR image $x_I$.

We apply the first three residual blocks of ResNet-18 \cite{He} as the SAR image feature extractor, denoted as $\mathcal{F}_{img}(x_I)$. Since the weak relationship between physics-specific topics $y_{phy}$ and image-specific features, we design the physics mapping layer (PML) denoted as $\mathcal{F}_{pml}$ to narrow the knowledge gap. The PML is composed of a convolution module and a fully-connected layer, mapping the image representations to physics topic space, denoted as $\phi_I=\mathcal{F}_{pml}(\mathcal{F}_{img}(x_I))$ where $\phi_I \in \mathbb{R}^K$.

The following objective function describes the soft semantic relations between them, that
\begin{equation}
\label{equ:soft}
L_{sr}(y_{phy}, \phi_I) = - \sum_{k \in \bm{K}_{vis}} y_{phy}^{k} \cdot \phi_I,
\end{equation}
where $\bm{K}_{vis}$ denotes the topics that can be represented by features from SAR vision domain. Equa. (\ref{equ:soft}) is a relaxed constraint, where only the related semantics are considered to be similar. As a comparison, the hard constraint is
\begin{equation}
\label{equ:hard}
L_{sr}(y_{phy}, \phi_I) = - \sum_{k \in \bm{K}_{vis}} y_{phy}^{k} \cdot \phi_I + \sum_{k \notin \bm{K}_{vis}} y_{phy}^{k} \cdot \phi_I,
\end{equation}
where the unrelated semantics ($k \notin \bm{K}_{vis}$) are additionally required to be highly different.

We choose the soft constraint in our method considering the semantic gap between SAR physics knowledge and the visual perception of image data. The follow-up experiments will discuss the differences of soft and hard constraints.

In order to simplify the gradient descent optimization, we modify Equa. (\ref{equ:soft}) as
\begin{equation}
\label{equ:costfun}
L_{sr}(y_{phy}, \phi_I) = - \sum_{k=1}^K \left( y_{phy}^k \log \frac{e^{\phi_I^k}}{\sum_i e^{\phi_I^i}} \right) \cdot \delta_k,
\end{equation} 
where $\delta_k$ is the activation term with a value of 1 or 0. We choose the locations where $y_{phy}^k \geqslant \alpha$ as activated ones ($\delta_k=1$) with a probability of $p_a$, otherwise deactivated ($\delta_k=0$). The parameter $\alpha$ filters the remarkable attributes in $y_{phy}$, that is, only the significant semantic topics are considered as possibly related. The probability $p_a$ decides only part of the semantic topics are selected to be related.

\subsection{Physics Injected Network}

The physics injected network (PIN) is designed to inject the physics-aware features obtained from the unsupervised PGN into the traditional deep neural network. The injected features provide abundant prior information for the deep network training and are adapted to satisfy the classification task as far as possible.

To decide which layer is appropriate to be taken for the injected physics-aware features is crucial. In our work, we select the output of ResBlk-3 as the physics-aware features to be injected, that is, $F_{PA}=\mathcal{F}_{img}(x_I)$. The decision will be discussed in the following experiments.

We propose a simple injection strategy that adding the transformed physics-aware features to the mid- and high-level layers of the traditional classification network successively. The blue modules in Fig. \ref{fig_method} are the conventional ResNet-18 classification networks, denoted as $\mathcal{F}_{rs18}$. The transform layers, shown in the red module in Fig. \ref{fig_method}, are designed to convert the physics-aware features to the same size of the destination, denoted as $\mathcal{F}_{tr}$. The transform module is composed of a 1$\times$1 convolution layer and an upsample layer if needed, for channel and feature size transformation respectively. The final output of the physics injected neural network is written as $Inj(\mathcal{F}_{rs18}(x_I), \mathcal{F}_{tr}(F_{PA}))$.

In the classification case, the cross entropy softmax loss function is widely used. Here, we denote the CE loss as
\begin{equation}
\label{equ:ce}
L_{ce}(x_I, y_{tgt}) = -\sum_i y_{tgt}^{i} \log c_i,
\end{equation}
where $c_i=Inj(\mathcal{F}_{rs18}(x_I^i), \mathcal{F}_{tr}(\mathcal{F}_{img}(x_I^i)))$.

In order to ensure the physics-aware features to be more adaptive to the classification task, we add the small weighted soft constraint $L_{sr}$ in Equa. (\ref{equ:costfun}) as an regularization term and fine-tune the PGN slightly in supervised classification training. The total loss function is written as
\begin{equation}
\label{equ:reg}
L_{ce}(x_I, y_{tgt}) + \lambda L_{sr}(y_{phy}, \phi_I).
\end{equation}

\section{Experiments}
\label{sec:exp}

In this section, we firstly introduce the hybrid Image-Physics data format and the experimented datasets. Then, we conduct several experiments to prove the effectiveness of our proposed method with sufficient discussions.

\subsection{Dataset and Experimental Setup}
\label{sec:data}

Most SAR image classification datasets, like OpenSARUrban \cite{Zhao2020}, only provide the processed SAR amplitude images $x_I$ for a better visual understanding. For the purpose of leveraging the underlying physics knowledge in SAR and meanwhile preventing large storage space for complex data $x$, we propose the hybrid Image-Physics (Img-Phy) data format to integrate $x_I$ and $\mathcal{L}(x)$ in a concise way, to accomplish the proposed PGIL method.

\begin{figure*}[!t]
\centering
\includegraphics[width=15cm]{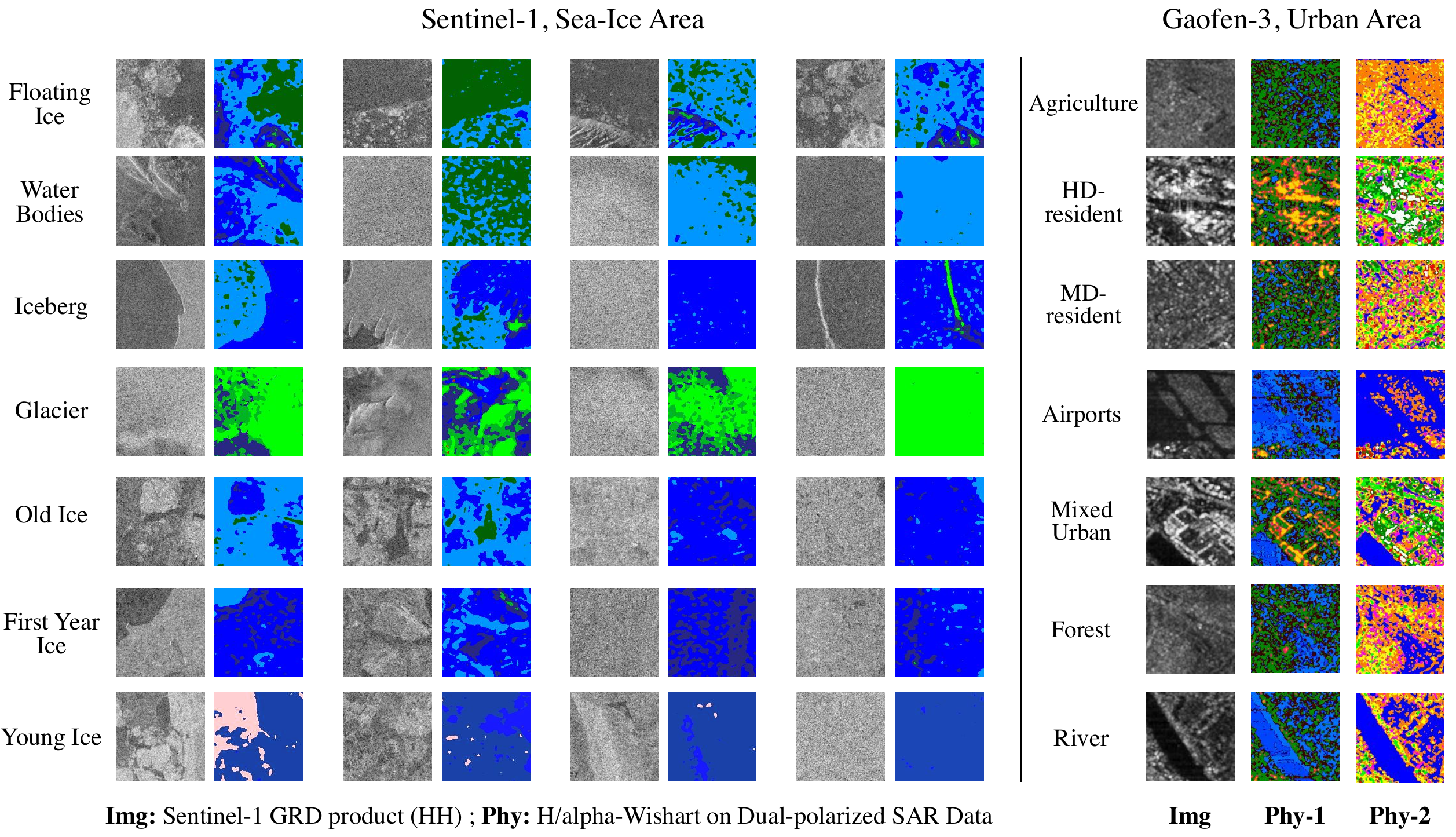}
\caption{The hybrid Img-Phy data examples. Left: sea-ice classification dataset with manual annotations. The Img part is the Sentinel-1 grounded range detected (GRD) product and the Phy part is the H/$\alpha$-Wishart result from the single-look complex (SLC) product. Right: the Gaofen-3 SAR image patches in urban areas. Phy-1: the HDEC-TFA result \cite{Huang2020}, Phy-2: the H/$\alpha$-Wishart result \cite{lee}, which are also given in Fig. \ref{fig_Lxexp}.}
\label{fig_data}
\end{figure*}

We mainly evaluate our method on a sea-ice classification dataset acquired by Sentinel-1, as shown in Fig. \ref{fig_data} (Left). The Sentinel-1 Interferometric Wide (IW) SAR data in polar region is downloaded \footnote{\url{https://scihub.copernicus.eu/}}, both single-look complex (SLC) and multi-looked Ground Range Detected High resolution (GRDH) product (HH channel). Seven sea-ice types are annotated with the patch size of 256$\times$256 for GRDH image, serving as $x_I$. Besides, the dual-polarized SLC data is processed by SNAP software \footnote{\url{https://step.esa.int/main/download/snap-download/}} to obtain the H/$\alpha$ labels, serving as $\mathcal{L}(x)$. To ensure $\mathcal{L}(x)$ almost covers the same area with $x_I$, some essential operations like multi-looking, grounded range projection, are required. Since the different pixel spacing between the processed SLC and GRDH, the Phy data is not square anymore, which is stored as a matrix with the size of about 187$\times$139. For better visualization, Fig. \ref{fig_data} shows the resized square Phy patch in RGB format, where each color represents a scattering label.

For a better evaluation, especially in case of limited labeled data, we randomly select 45, 35, 25, 15, and 5 samples from each class for training, denoted as train-45, train-35, train-25, train-15, and train-5, respectively. The test set is fixed, as shown in Table \ref{tab:ice}.

 \begin{table}
 \caption{The sea-ice classification dataset of Sentinel-1.}
 \centering
 \label{tab:ice}
 \begin{tabular}{cccccccc}
 \toprule
 class   & YI & GL & FY & WT & IB & FI & OI \\
 \midrule
 train   & \multicolumn{7}{c}{train-45,  train-35,  train-25,  train-15,  train-5}\\
 test    & 1399 & 51  & 129 & 406 & 154 & 76 & 1107 \\
 \bottomrule
 \end{tabular}
 \end{table}

In addition, a Gaofen-3 SAR scene image covering a wide urban area is experimented \footnote{\url{https://www.ietr.fr/GF3/}}, shown in Fig. \ref{fig_data} (Right). Seven land cover and land use classes are annotated. The training/test details are listed in Table \ref{tab:urban}. We present two different types of $\mathcal{L}(x)$ introduced before, that are Phy-1 (the HDEC-TFA result \cite{Huang2020} on single-polarized HH channel data) and Phy-2 (the H/$\alpha$-Wishart result \cite{lee} on full-polarized data). In the following experiments, we will discuss the physics guided learning results with different Phy data.

 \begin{table}
 \caption{The land-use and land-cover classification dataset of Gaofen-3.}
 \centering
 \label{tab:urban}
 \begin{tabular}{cccccccc}
 \toprule
 class   & WT & AL & MF & AP & HD & MD & MU \\
 \midrule
 train   & 20 & 20 & 20 & 10 & 20 & 20 & 20 \\
 test    & 157 & 80  & 650 & 34 & 231 & 871 & 775 \\
 \bottomrule
 \end{tabular}
 \end{table}

When using H/$\alpha$-Wishart result of the polarimetric SAR as the Phy data, the obtained $\mathcal{L}(x)$ is with $N_s=9$ classes of physical scattering characteristics, corresponding to nine zones in H/$\alpha$ plane shown in Fig. \ref{fig_bkg}(a). While for HDEC-TFA result of the single channel (HH) SAR image, $N_s$ is set to 15 according to \cite{Huang2020}. In the topic modeling for physics guided signals generation, the vocabulary size $N_v$ and the topic number $K$ of the LDA model are set to 500 and 175, respectively. Note that the topic number $K$ is a critical parameter in the algorithm, that will be under discussions in the following experiments. To determine $\delta_k$ in Equation (\ref{equ:costfun}), we set $\alpha$ and $p_a$ to 0.1 and 0.9, respectively. The following discussions will illustrate the strategy of parameter setting. 

The physics guided learning is optimized by stochastic gradient descent (SGD) with a fixed learning rate of 0.05, and the momentum is set to 0.9 by default. All Img-Phy pairs in the dataset are fed into the PGN for training, lasting 200 epochs in total. The physics injected learning only takes annotated data to train $\mathcal{F}_{rs18}$ and $\mathcal{F}_{tr}$. The initial learning rate is set to 0.001 and the cosine annealing strategy is applied to decrease the learning rate to $10^{-8}$ in the last 3 epochs of 50 in total. The soft constraint regularization term in Equa. (\ref{equ:reg}) is weighted by $\lambda$ set to 0.1.

All experiments are conducted on a workstation of 64 bit Linux operating system, with 64G RAM and NVIDIA RTX 3090 graphics card of 24GB GDDR6X VRAM clocked at 1700 MHz.

\subsection{Unsupervised Physics Guided Learning}

The PGN learns physics-aware features $F_{PA}$ from all Img-Phy pairs. To evaluate the discriminative ability of $F_{PA}$ in semantic domain, we train a support vector machine (SVM) on $F_{PA}$ to predict $y_{tgt}$. In this section, we will firstly discuss how the topic number $K$, and the activation strategy $\delta_k$ effect the discriminative physics-aware feature learning based on the sea-ice classification dataset of Sentinel-1. Then, the characteristics of physics-aware features, as well as the differences between hard and soft constraint, are analyzed. At last, we additionally evaluate the effectiveness of the proposed physics guided learning approach on Gaofen-3 SAR data covering a wide urban area.

\subsubsection{Hyperparameter Discussion}

Firstly, we set $\alpha$ and $p_a$ to 0.1 and 0.9, respectively, and discuss the topic number $K$. The classification results are shown in Table \ref{tab:topicnum}, where the highest overall accuracy or F1-score are marked in red. We check out six different values of $K$ (25, 50, 100, 150, 175, 200) and train the SVM classifier on 5 different training sets. It can be observed from Table \ref{tab:topicnum} that a larger $K$ almost leads to a better result. Since the fully-connected layer in the PML module is determined by the topic number, a large $K$ would introduce plenty of parameters and increase the computation load. Consequently, we decide $K=175$ for a better trade-off.

\begin{table*}
\caption{The SVM performance (Overall Accuracy / F1-score ($\%$)) of physics-aware features in case of different topic numbers. Parameter setting: $\alpha=0.1, p_a=1.0$}
\label{tab:topicnum}
\centering
\begin{tabular}{cccccc}
\toprule
\midrule
$K$	& train-45	& train-35	& train-25	& train-15	& train-5 \\
\midrule
25 & 77.57 / 69.34	& 77.21 / 69.34 & 77.24 / 68.85  & 77.33 / 68.85	& 74.14 / 66.64 \\
50 & 81.88 / 75.98	& 80.70 / 74.29 & 80.85 / 74.66  & 78.51 / 72.05 	& 78.00 / 70.78 \\
100 & 83.20 / 78.54	& 81.88 / 76.26	& 81.88 / 75.52	 & 80.34 / 74.00 	& {\color{red}78.45} / 71.46	 \\
150 & 84.14 / 78.79	& 83.53 / 78.39 & 82.33 / 76.28  & 82.00 / 75.47 	& 77.81 / 70.86 \\
\textbf{{\color{red}175}} & {\color{red}84.18 / 79.24} & {\color{red}83.62 / 78.51} & 82.03 / {\color{red}76.95}  & 81.19 / 75.05 	& 78.39 / 71.53  \\
200 & 82.66 / 77.20 & 82.51 / 76.99 & {\color{red}82.45} / 76.78  & {\color{red}82.03 / 76.20} 	& 78.12 / {\color{red}72.29}   \\
\bottomrule
\end{tabular}
\end{table*}

\begin{figure}[!t]
\centering
\includegraphics[width=8cm]{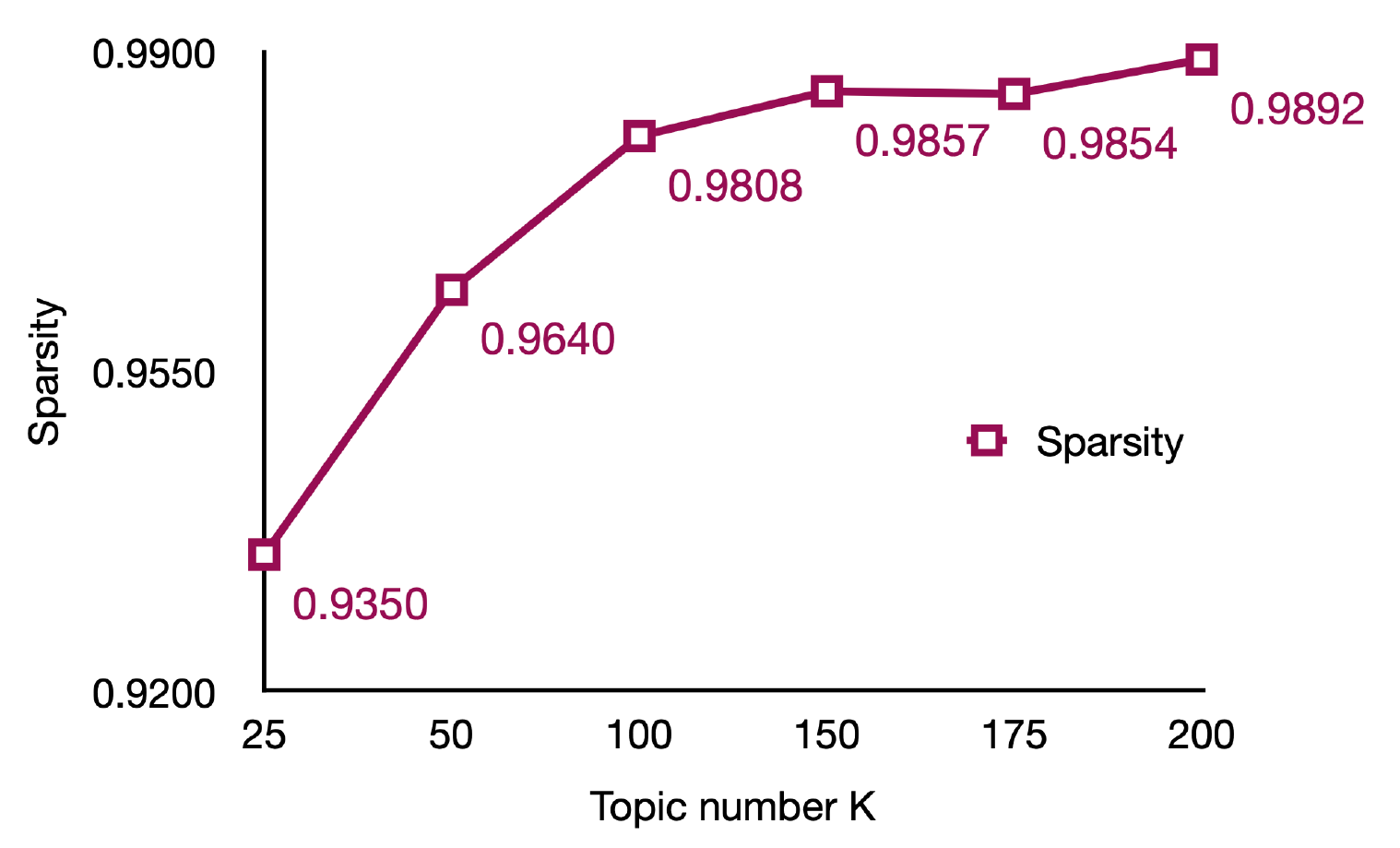}
\caption{The topic sparsity of LDA model in case of different topic number $K$.}
\label{fig_sparsity}
\end{figure}

The topic number $K$ decides the shared attributes space $\mathbb{R}^K$ where the image representation is mapped. The assumption is proposed to build a bridge between $\phi_I \in \mathbb{R}^K$ and $\{\varphi_k\}$. A larger $K$ ensures more fine-grained physics attributes, so that the soft constraint in Equa. (\ref{equ:soft}) can be more precise. We calculate the sparsity of BoT vector $\bm{\varphi}$, defined as
\begin{equation}
sparsity = 1 - \frac{||\bm{\varphi}||_0}{K},
\end{equation}
where $||\cdot||_0$ denotes the L0 norm. Fig. \ref{fig_sparsity} plots the sparsity of BoT representation in case of different topic numbers. We find the fine-grained physics attributes lead to a more sparse representation of BoT encoding. The BoT sparsity is also highly relate to the physics-aware feature performance. Intuitively, a sparsity greater than 0.985 can be regarded as a good choice with the topic number $K$ no less than 150.

Next, we will discuss how to fix the activated physics attributes $\delta_k$ in Equa. (\ref{equ:costfun}). $\delta_k$ is determined by $\alpha$ and $p_a$, where $\alpha$ filters the prominent attributes in $y_{phy}$ as the candidates, and $p_a$ randomly select the potential attributes to calculate the constraint. Table \ref{tab:alpha} shows the SVM classification results of different $\alpha$, with $K=175$ and $p_a=0.9$. When $\alpha$ equals 0, some unconsidered attributes ($\phi_k$ close to 0) may be included to guide the network to learn the insignificant features. It is better to set $\alpha$ as a small value but greater than 0, e.g. 0.1 in our case. 

\begin{table*}
\centering
\caption{The SVM performance (Overall Accuracy / F1-score ($\%$)) of physics-aware features in case of different $\alpha$. Parameter setting: $p_a=0.9, K=175$}
\label{tab:alpha}
\begin{tabular}{cccccc}
\toprule
\midrule
$\alpha$ & train-45  & train-35  & train-25  & train-15  & train-5 \\
\midrule
0 & 84.05 / 79.15 & 83.32 / 78.05 & 82.69 / 77.32  & 81.55 / 75.55  & 78.20 / 71.46 \\
\textbf{{\color{red}0.1}} & {\color{red}84.18 / 79.24} & {\color{red}83.62 / 78.51} & 82.03 / 76.95  & 81.19 / 75.05   & {\color{red}78.39 / 71.53}  \\
0.2 & 83.96 / 78.98 & 83.20 / 77.93 & {\color{red}82.90 / 77.54}  & {\color{red}82.42 / 76.43}   & 78.09 / 71.44 \\
0.3 & 81.58 / 76.11 & 82.21 / 76.55 & 81.55 / 75.69  & 79.38 / 73.20  & 75.89 / 70.32  \\
\bottomrule
\end{tabular}
\end{table*}

Table \ref{tab:pa} discusses the values of $p_a$ in the context of $\alpha=0.1$ and $K=100$. $p_a$ is the probability for randomly selecting the potential attributes in $y_{phy}$. The value of $p_a$ from 0.5 to 1.0 indicates the constraint becomes rigid. The result shows it is better to choose a greater values of $\alpha$, which demonstrates a majority of remarkable attributes should be considered. Here, we choose $\alpha=0.9$ in our experiments.

To summarize, the topic number $K$ is the most important hyperparameter in PGN learning which can be determined by the sparsity of BoT representation. $\delta_k$ controls a relax activation of physics attributes, decided by a relatively casual value of $\alpha$ and $p_a$. We recommend to set $\alpha$ to 0.1 or 0.2, and $p_a$ to 0.9 or 0.8, respectively.

\begin{table*}
\caption{The SVM performance (Overall Accuracy / F1-score ($\%$)) of physics-aware features in case of different $p_a$. Parameter setting: $\alpha=0.1, K=100$}
\label{tab:pa}
\centering
\begin{tabular}{cccccc}
\toprule
\midrule
$p_a$	& train-45	& train-35	& train-25	& train-15	& train-5 \\
\midrule
1.0 & 83.20 / 78.54	& 81.88 / 76.26 & 81.88 / 75.52  & 80.34 / 74.00	& 78.45 / 71.46 \\
\textbf{{\color{red}0.9}} & 83.35 / 78.09	& 83.02 / 77.37 & {\color{red}82.48} / 76.00  & {\color{red}81.13 / 74.74} 	& {\color{red}79.23 / 71.87} \\
0.8 & {\color{red}83.87 / 78.73}	& {\color{red}82.93} / 77.51	& 82.45 / {\color{red}76.64}	 & 81.10 / 74.60 	& 78.24 / 71.84	 \\
0.7 & 83.17 / 77.93	& {\color{red}82.93 / 77.66} & 81.58 / 76.30  & 79.74 / 73.70 	& 78.27 / 71.83 \\
0.6 & 82.18 / 76.81 & 81.52 / 76.54 & 80.67 / 75.25  & 79.74 / 73.65 	& 77.39 / 71.32  \\
0.5 & 80.34 / 74.39 & 80.79 / 74.71 & 79.86 / 73.17  & 79.02 / 72.58 	& 75.80 / 68.96   \\
\bottomrule
\end{tabular}
\end{table*}

\subsubsection{The Physics-Aware Features Discussion}

The PGN is comprised of the ResNet-13 backbone and the 2-layer PML module. The ResNet-13 backbone $\mathcal{F}_{img}$ extracting hierarchical features from $x_I$ is basically image specific, with the higher-level features becoming closer to semantic meaning. The PML module transforms $\mathcal{F}_{img}(x)$ to the physics attributes space $\mathbb{R}^K$, building the semantic relation between image representation and physics knowledge. The physics-aware features are expected to be discriminative in the classification semantic domain and also with physics awareness. 

\begin{figure*}
\centering
\includegraphics[width=17cm]{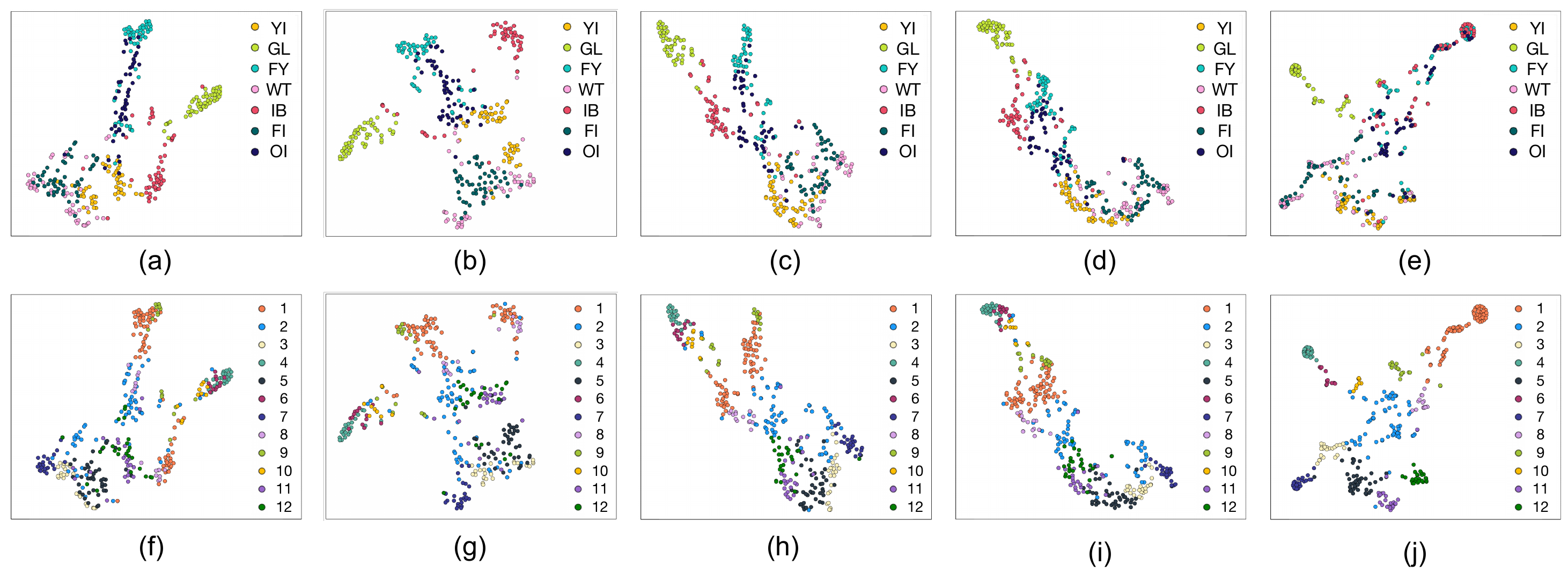}
\caption{The feature visualization of training data in different layers. Features in (a)(f), (b)(g), (c)(h), (d)(i), are from ResBlk-2, ResBlk-3, PML-conv, and PML-fc, respectively. (e)(j) are BoT representations. (a)(b)(c)(d)(e) are marked with true labels. (f)(g)(h)(i)(j) are marked with $k$-means cluster labels of BoT representations.}
\label{fig_featvis}
\end{figure*}

We analyze the outputs of different layers in PGN to demonstrate the semantic discrimination and the physics awareness of features. The SVM classification results are used to indicate the semantic discrimination, as shown in the first row in Table \ref{tab:pafeatsvm}. The features from ResBlk-3 reach the highest classification accuracy of 84.18\%, followed by an overall accuracy of 82.39\% for the features from the convolution layer in PML module. The results demonstrate how the feature discrimination in semantic level changes with the physics guided neural network. Note that the physics BoT encoding only achieves 61.53\% in classification, which indicates $y_{phy}$ is highly physics specific and the semantic gap is truly existed between $y_{phy}$ and $y_{tgt}$. Even so, $y_{phy}$ can guide the PGN to learn the discriminative features close to $y_{tgt}$ successfully, with the designed objective function. Intuitively, the first row in Fig. \ref{fig_featvis} displays the annotated labels with different colors, indicating the feature discrimination in semantic level. We can observe that the BoT encodings are confused in understanding the semantic labels, since the SAR images in the same class may have different physics attributes. After the physics guided learning, the feature discrimination in semantic level has been improved from PML-fc to ResBlk-3 layer. Due to the lower level of ResBlk-2, the features in Fig. \ref{fig_featvis}(a) are not as discriminative as those of ResBlk-3 in Fig. \ref{fig_featvis}(b).

\begin{table*}
\caption{The SVM classification results (OA / F1-score), and the physics awareness analysis of features (quantified by silhouette score) from different layers in PGN. The soft and hard constraints in Equa. (\ref{equ:soft}) and (\ref{equ:hard}) are both explored.}
\label{tab:pafeatsvm}
\centering
\begin{tabular}{ccccccc}
\toprule
\midrule
$F_{PA}$ & Constraint	& ResBlk-2	& ResBlk-3	& PML-conv & PML-fc	& BoT \\
\midrule
\multirow{2}{*}{\makecell[c]{\textbf{SVM result} \\ OA / F1-score ($\%$)}} & soft &	79.05 / 72.75		&	{\color{red}84.18 / 79.24}		&	{\color{blue}82.39 / 76.84}		&	77.90 / 71.74 & 61.53 / 51.86	 \\
& hard & 78.45 / 71.55 & 82.99 / 78.05 & 81.85 / 75.90 & 77.45 / 71.40 & 61.53 / 51.86  \\
\midrule
\multirow{2}{*}{\makecell[c]{\textbf{Physics Awareness} \\ Silhouette score}} & soft & 0.0157 & 0.0351 & 0.0904 & 0.1087 & {\color{red}0.4766} \\
& hard & 0.0181 & 0.0380 & 0.0949 & 0.1202 & 0.4766 \\
\bottomrule
\end{tabular}
\end{table*}

Additionally, we demonstrate the physics awareness of features by visualization and quantitative metrics, shown in the second row of Fig. \ref{fig_featvis} and Table \ref{tab:pafeatsvm}, respectively. Given the BoT encodings of physics attributes, we apply a $k$-means algorithm to cluster the BoT representations into $k$ classes as the color identification in the second row of Fig. \ref{fig_featvis}. Thus, (f)(g)(h)(i)(j) indicate the feature discrimination in physics level, that is, the samples with the same color have similar physics attributes. We can observe how the features become physics aware with the help of PML module. Table \ref{tab:pafeatsvm} also lists the silhouette coefficient of features which reflect the separation between clusters. The silhouette coefficient values between -1 and 1, and a higher silhouette score indicates better discrimination of physics information in this feature space, that is, stronger physics awareness of features. It gradually decreases from physics topic space to image feature space, but still keep a positive value of 0.0351 in ResBlk-3. As a comparison, the silhouette score of features in ResBlk-3 of the traditional CNN learning model is -0.0085, indicating physics unawareness. Hence, we assert the PGN is able to learn the physics-aware features.


\subsubsection{Hard and Soft Constraint Discussion}

We discuss the soft and hard constraint objective functions defined in Equa. (\ref{equ:soft}) and (\ref{equ:hard}). The soft constraint only emphasizes the common semantics from physics and vision domain to be highly similar, while the hard constraint additionally restricts the specific ones to be different. The SVM classification results in Table \ref{tab:pafeatsvm} indicate that the soft constraint guides the PGN to learn more general features which achieve better performance in classification. The listed silhouette coefficients of hard constraint in Table \ref{tab:pafeatsvm} are larger than those of soft constraint, which demonstrate the the learned features by hard constraint are more physics specific but less semantic discriminative. The discussion explains the semantic gap between the physics attributes and the image features of SAR, encouraging us to find a trade-off in learning the physics-aware features.

\subsubsection{Generalization Analysis of PGN}

We additionally take the other Gaofen-3 SAR data, as shown in Fig. \ref{fig_data} (right), to demonstrate the effectiveness and generalization ability of the proposed PGN, besides the polarimetric characteristics derived from H/$\alpha$-Wishart method \cite{lee1999}. In our previous work \cite{Huang2020}, we verified that the HDEC-TFA method could automatically discover the time-frequency properties of SAR target in high resolution SAR images, especially for some man-made targets with characteristic scattering behaviors. It is based on the physics meanings of time-frequency analysis on complex SAR data, which reveals the scattering variation on different azimuth angles and range bandwidths. We apply different physics information on PGN to illustrate our proposed method can be integrated with various physical models. 

\begin{table}
\caption{The SVM classification results of features in physics guided learning with different physics information, compared with the supervised end-to-end CNN training. (Overall Accuracy / F1-score ($\%$))}
\label{tab:gf3svm}
\centering
\scriptsize
\begin{tabular}{cccc}
\toprule
\midrule
Phy Data  & None (CNN)  & HDEC-TFA (Phy-1)  & H/$\alpha$-Wishart (Phy-2) \\
\midrule
Result &  52.43 / 48.51   & {\color{red}68.76 / 62.32} &  62.54 / 58.71   \\
\bottomrule
\end{tabular}
\end{table}

The SAR image in urban city with dense buildings and man-made targets is more complicated than polar area. With very limited annotation, it is difficult for supervised CNN training to learn generalized and discriminative features. Table \ref{tab:gf3svm} shows the CNN only achieves an accuracy and F1-score of 52.43\% and 48.51\%, respectively, with severe overfitting. The SVM classification is utilized to evaluate the physics-aware features by PGN with different Phy data. As recorded in Table \ref{tab:gf3svm}, the classification result of physics-aware features guided by Phy-1 (HDEC-TFA) signals improves 16.33\% in accuracy than CNN training result, 6.22\% better than guided by H/$\alpha$-Wishart scattering characteristics. It also verifies the effectiveness of HDEC-TFA learning approach to extract significant physics properties under circumstance of no polarimetric information available.

\begin{figure*}[!t]
\centering
\includegraphics[width=15cm]{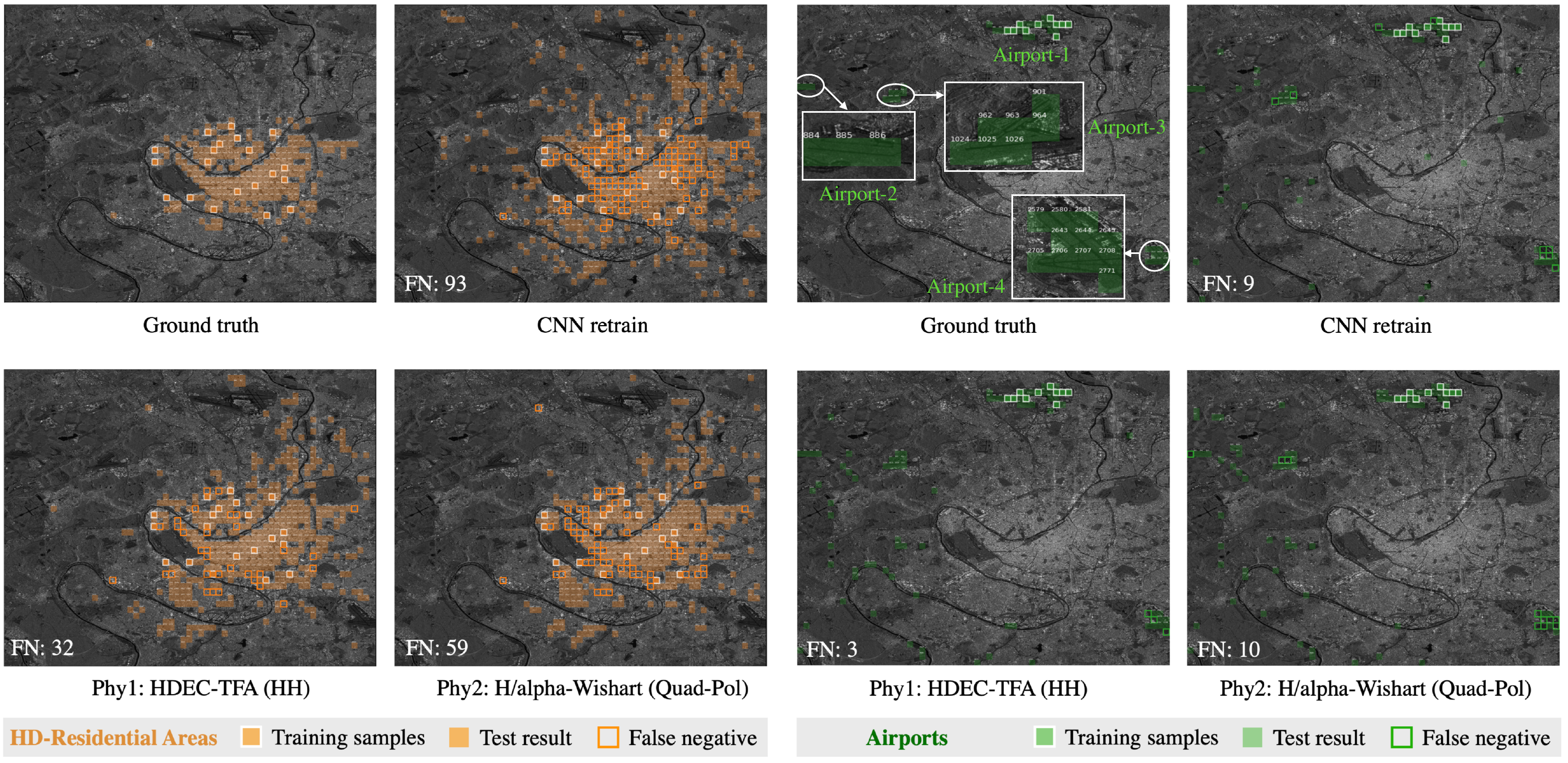}
\caption{The physics guided learning result with different physics information, compared with the CNN result on Gaofen-3 SAR image.}
\label{fig_gf3vis}
\end{figure*}

Fig. \ref{fig_gf3vis} visualizes the annotated ground truth, the test results of CNN, PGN with Phy-1 and Phy-2 on two classes, \textit{High-density residential area} and \textit{Airport}. The scattering in high-density residential area is strong and extremely complicated so that the visual interpretation is difficult. However, the physical characteristics in this area, as shown in Fig. \ref{fig_Lxexp}, are more distinctive than other regions, especially in HDEC-TFA case. Consequently, PGN with Phy-1 (HDEC-TFA) achieves the best prediction result on high-density residential area with minimum false negative samples. There are 4 airports in the SAR image, as shown in Fig. \ref{fig_gf3vis}, where only 10 annotated samples in Airport-1 are used for training. The PGN with Phy-1 can remarkably predict the remaining 3 unseen airports with only 3 false negative samples, which is much superior than traditional data-driven CNN model.

\subsection{Supervised Physics Injected Learning}

In this section, we discuss the effectiveness of PIN module on sea-ice classification dataset in Table \ref{tab:ice} (train-45 as the training data). The physics-aware representation $F_{PA}$ is considered as an off-the-shelf feature from PGN to be injected. The baseline is the traditional CNN method, that is, training $\mathcal{F}_{rs18}$ with only labeled amplitude SAR images. We both test the training from the scratch strategy and the transfer learning strategy using a SAR image pre-trained model proposed in \cite{Huang2021}. The classification results are shown in Table \ref{tab:abs}, where the retraining accuracy is only 74.55\% and the transfer learning result is 82.57\%. It shows that the limited labeled data is insufficient to train a very deep neural network from scratch. After the physics-aware features injection, however, the retraining results improve more than 10\%, as shown in Table \ref{tab:abs}.

\begin{table}
  \caption{The ablation study of physics injected learning. (OA ($\%$))}  
  \label{tab:abs}
  \centering
  \begin{tabular}{ccccc}
    \toprule
    \textbf{inj-2} & \textbf{inj-3}  & \textbf{inj-4} & retrain (\%) & pre-trained (\%) \\
    \midrule
    - & - & - & 74.55 $\pm$ 1.05 & 82.57 $\pm$ 0.82 \\
    $\surd$ & - & - & 82.56 $\pm$ 2.36  & 81.56 $\pm$ 1.89  \\
    - & $\surd$ & - & {\color{blue}84.52 $\pm$ 0.39}  & {\color{blue}85.19 $\pm$ 0.75}  \\
    - & - & $\surd$ & 82.15 $\pm$ 1.18  & 84.05 $\pm$ 0.89  \\
    $\surd$ & $\surd$ & - & 82.11 $\pm$ 0.31  & 84.15 $\pm$ 0.83  \\
    - & $\surd$ & $\surd$ & 82.77 $\pm$ 1.47  & 83.73 $\pm$ 1.35  \\
    $\surd$ & $\surd$ & $\surd$ & {\color{red}84.96 $\pm$ 0.36} & {\color{red}86.40 $\pm$ 0.35} \\
    \bottomrule
  \end{tabular}
\end{table}

We discuss the different locations where the physics-aware features are injected, including the ResBlk-2, ResBlk-3, and ResBlk-4, denoted as inj-2, inj-3, and inj-4 in Table \ref{tab:abs}, respectively. With the same depth of $F_{PA}$ and the features of ResBlk-3, the injection in ResBlk-3 reaches the best performance of single-injection strategy, marked in blue. The results indicate that the obtained physics-aware feature is with abstract meanings, but still has semantic gap with the semantic features for target task. We also find that the multi-layer injection can improve the classification most in both retraining and transfer learning cases.

The unsupervised PGN training costs about 4.45h, with the batchsize of 300 and training epochs of 200. Afterwards, the supervised PIN only takes 13.25 minutes for training, with the batchsize of 100 and training epochs of 100.

\subsection{Ablation Study}

In order to demonstrate the effectiveness of each module in the proposed method, we conduct the detailed ablation study of each module on sea-ice dataset (train-45 as the training data). 

As shown in Table \ref{tab:abs_PGPI}, the baseline is set as retraining the CNN model from scratch, achieving an overall accuracy of 74.55\% with strong overfitting. As a contrast, the proposed method reaches 85.27\% in average, obtaining about 10.72\% improvements. We discuss the effectiveness of the following four parts:

\begin{enumerate}
  \item Explainable Models (XM): Based on the existing physical scattering labels, generating the abstract BoT encodings with LDA as the physics-guided signals. 
  \item Physics Guided Network (PGN): Learning the physics-aware features guided by physics BoT representations.
  \item Physics Injected Network (PIN): Retraining the CNN with injecting physics-aware features.
  \item Self-Adaptive Learning (SAL): Fine-tuning the physics guided network during the PIN training with a combined loss.
\end{enumerate}

Table \ref{tab:abs_PGPI} shows the ablation experiments, excluding each part above-mentioned separately. It is clear that the XM and PGN play the most important roles in the proposed method. Generating an appropriate physics guided signal $y_{phy}$ in XM remarkably effects the quality of injected features learned by PGN, with the accuracy increasing from 80.61\% to 84.96\%. The PGN module ensures the injected knowledge more related to the target task. Otherwise, directly injecting the BoT representation in Equa. (\ref{equ:bot}) only has an accuracy of 78.78\% in average, and the proposed PGN contributes 6.18\% improvement. The PIN learning makes further efforts on fusing the physics knowledge and vision features together, which has an improvement of about 2.37\%. The SAL part which makes the physics-aware features more adaptive to the target task slightly improves the result. 

\begin{table}
  \caption{The ablation study of different modules. (OA ($\%$))}  
  \label{tab:abs_PGPI}
  \centering
  \scriptsize
  \begin{tabular}{cccccc}
    \toprule
    \textbf{XM} & \textbf{PGN}  & \textbf{PIN} & \textbf{SAL} & description & result (\%) \\
    \midrule
    \multicolumn{4}{c}{Baseline}	&	retrain CNN	&	74.55 $\pm$ 1.05	\\
    \midrule
    - & $\surd$ & $\surd$ & - & use $\mathcal{L}(x)$ as $y_{phy}$ & 80.61 $\pm$ 1.18  \\
    $\surd$ & - & $\surd$ & - & inject Equa. (5) & 78.78 $\pm$ 1.67  \\
    $\surd$ & $\surd$ & - & - & fine-tune PGN  & 82.59 $\pm$ 1.47 \\

    $\surd$	&	$\surd$	&	$\surd$	& - &	optimizing Equa. (\ref{equ:ce})	&	84.96 $\pm$ 0.36 	\\
    
    $\surd$ & $\surd$ & $\surd$ & $\surd$ & optimizing Equa. (\ref{equ:reg}) & {\color{red}85.27 $\pm$ 0.28}  \\
    \bottomrule
  \end{tabular}
\end{table}

Additionally, we compare some self-supervised learning methods in computer vision field \cite{Wu_2018_CVPR,pmlr-v119-chen20j} and also for PolSAR data \cite{Ren2021} with the proposed PGIL, since they all establish pretext tasks for unsupervised learning the feature embeddings. NPID \cite{Wu_2018_CVPR} learned the optimal feature via instance-level discrimination, while SimCLR \cite{pmlr-v119-chen20j} conducted the contrastive learning based on data-augmentation, both focusing on image contents. MI-SSL \cite{Ren2021} was proposed for PolSAR land cover classification, learning discriminative high-level features between multi-modal representations of PolSAR data. In order to adapt MI-SSL method to our case, we changed the SSL input of multi-modal features for full-polarized SAR data to our Img-Phy pairs. The results are listed in Table \ref{tab:self}. Although NPID and SimCLR perform well in natural image classification, such as ImageNet, the results demonstrate the pretext tasks of instance-level discrimination and contrastive learning via data-augmentation almost fail in SAR image classification. MI-SSL performs better for comparison, because it considers the multiple representations of SAR data. Our proposed PGIL reaches the best among them.

\begin{table}
  \caption{The comparison with self-supervised learning methods on sea-ice classification dataset. (OA ($\%$))}
  \label{tab:self}
  \centering
  \small
  \begin{tabular}{ccc}
    \toprule
    Method & Description & result (\%) \\
    \midrule
    \makecell{NPID \\ \cite{Wu_2018_CVPR}} & \makecell{instance-level \\ discrimination} & 66.56 $\pm$ 2.95 \\ 
    \midrule
    \makecell{SimCLR \\ \cite{pmlr-v119-chen20j}} & \makecell{data-augmented \\ contrastive learning} & 74.02 $\pm$ 0.78 \\
    \midrule
    \makecell{MI-SSL \\ \cite{Ren2021}} & \makecell{self-supervised \\ for PolSAR data} & 77.75 $\pm$ 1.41 \\
    \midrule
	PGIL	& Proposed	& 85.27 $\pm$ 0.28 	\\
    \bottomrule
  \end{tabular}
\end{table}

\begin{figure*}
\centering
\includegraphics[width=1.0\textwidth]{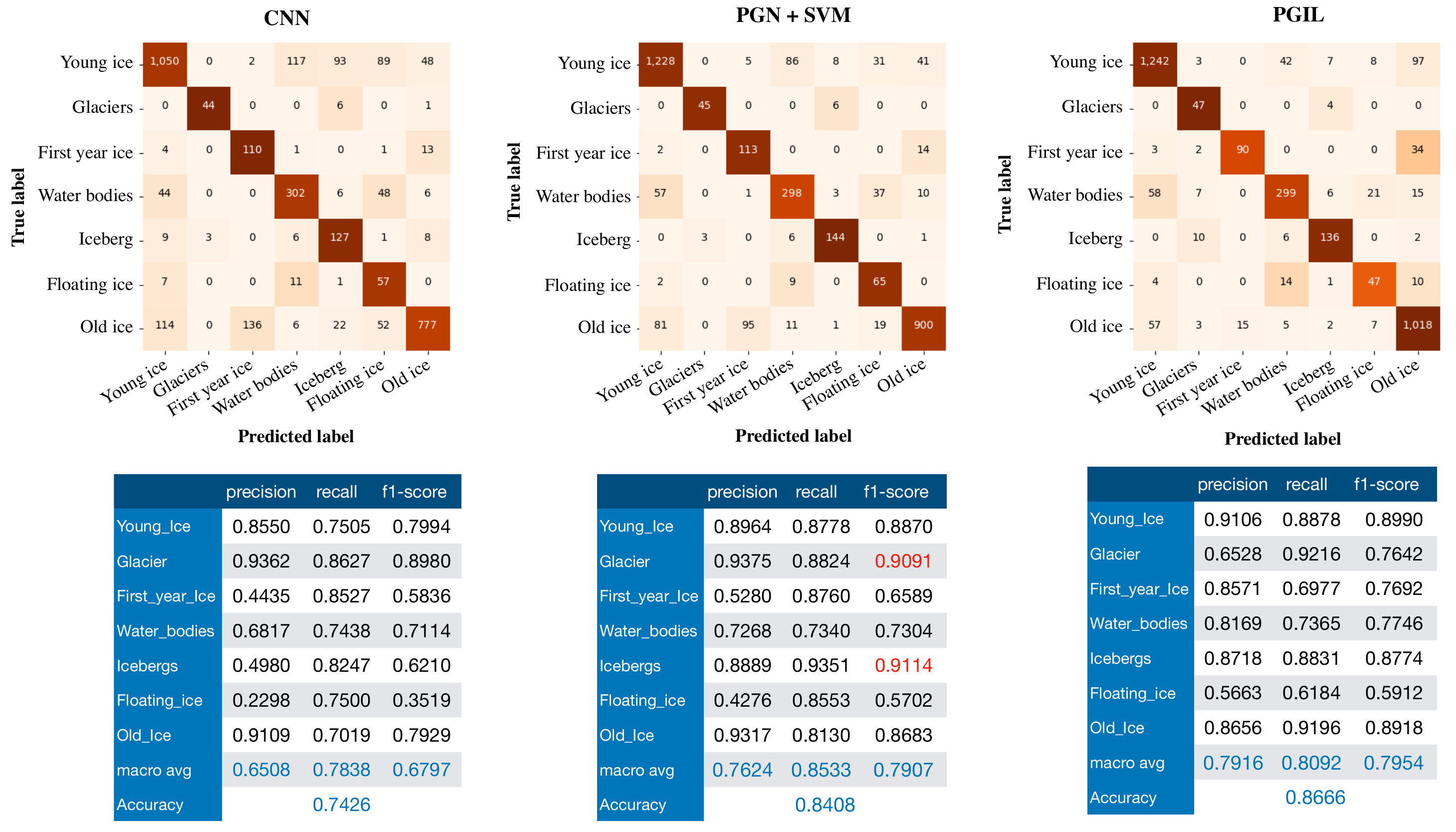}
\caption{The confusion matrix and the classification metrics (precision, recall, f1-score, and overall accuracy) of CNN learning, PGN + SVM, and PGIL performances are shown, respectively.}
\label{fig_cmatrix}
\end{figure*}

\subsection{Interpretability Discussion}

In this section, we use the sea-ice classification case to demonstrate the physics explainability of the proposed method. The explainability lies in the following two aspects. Firstly, the topic modeling for physics information provides explainable representations for each SAR image patch. Secondly, the PGN and PIN maintain the explainable physics consistency of features to learn reasonable results and prevent overfitting during automatic training.

\subsubsection{Physics Explanation of $y_{phy}$}

The PGN optimization is driven by the physics guided signals $y_{phy}$, denoted as $\mathcal{G}(\mathcal{L}(x))=\{\phi_1, \phi_2, ..., \phi_K\}$ in Equa. (5). The LDA topic modeling processing $\mathcal{G}$ explains the physical scattering characteristics $\mathcal{L}(x)$ by a combination of topics. Each latent topic is represented by a set of specific words, that is, the physical scattering characteristics distribution in a small area of SAR image (defined as "word" in LDA). The weight assigned ($\phi_i$) describes the probability of the SAR image patch belonging to the topic $i$. This benefits the understanding of hidden semantic structure between scattering labels of a large-scale SAR image area at an aggregate level. 


\begin{figure*}
\centering
\includegraphics[width=1.0\textwidth]{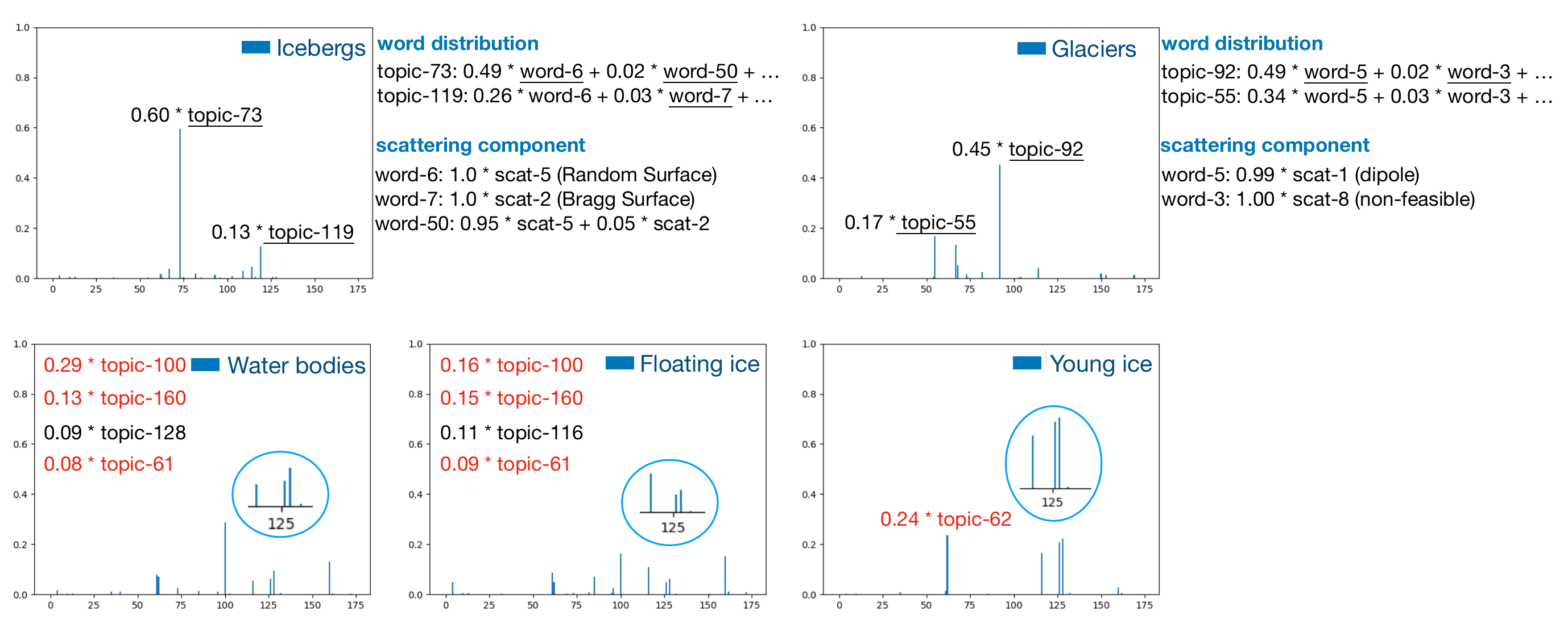}
\caption{The averaged physics BoT representation of training data for each class. The water bodies and the floating ice have very similar physics attributes.}
\label{fig_ldavis}
\end{figure*}

Fig. \ref{fig_ldavis} presents the averaged physics topic distribution of training data for some selected sea-ice classes, with topic number $K=175$. Since we have discussed previously that most SAR patches are with highly sparse physics topic representations, the classes whose distribution concentrated in fewer topics, such as iceberg and glaciers shown in Fig. \ref{fig_ldavis}, are more characteristic. The iceberg is mainly represented by topic-73 weighted 0.6 and topic-119 weighted 0.13. The word distribution of each topic is also given in Fig. \ref{fig_ldavis}, where each word can be explained by the physical scattering properties. In this sea-ice hybrid Phy-Img dataset, word-6 and word-7 are mostly random surface and Bragg surface scattering, respectively, and word-50 is a mixture of these two scattering properties.

Also, we list the four dominated topics and the topic weights of water bodies and floating ice in Fig. \ref{fig_ldavis}. The result indicates the physics attributes of water bodies and the floating ice are similar in semantic level. The inference can be proved from the semantic definition of floating ice -- any form of ice found floating in water \footnote{\href{https://www.canada.ca/en/environment-climate-change/services/weather-manuals-documentation/manice-manual-of-ice/chapter-1.html\#Floating_Ice}{The semantic definition of floating ice}}, that is to say, the SAR image patch with floating ice probably includes water bodies. Additionally, there are similar topics combinations in water bodies, floating ice, and young ice, shown in the zoom-in region of Fig. \ref{fig_ldavis}. The young ice has another specific topic of 62, indicating this class could have two kinds of representative physics attributes. 

\subsubsection{Explanation of PGIL Results}

Since the PGN is trained with the explainable BoT physics encoding, the above inferred information can explain the discrimination of physics-aware features in Fig. \ref{fig_featvis}(b). With more characteristic topic distribution of glaciers and icebergs, their physics-aware features are the most discriminative, and the SVM classification results show the two classes achieve the best F1-score of 0.909 and 0.911, respectively, as shown in Fig. \ref{fig_cmatrix}. In addition, the previous analysis explains the similar feature distribution of water bodies and floating ice shown in Fig. \ref{fig_featvis}(b), and the features of young ice have two different characteristics, one of which is close to water bodies and floating ice. As a result, we can observe in the confusion matrix that the most misclassified test samples in water bodies are predicted as young ice and floating ice, with a number of 57 and 37, respectively. Among the 11 false negative samples in the floating ice class, 9 are classified to water bodies and 2 to young ice.  

\begin{figure*}
     \centering
     \begin{subfigure}[b]{0.95\textwidth}
         \centering
         \includegraphics[width=0.9\textwidth]{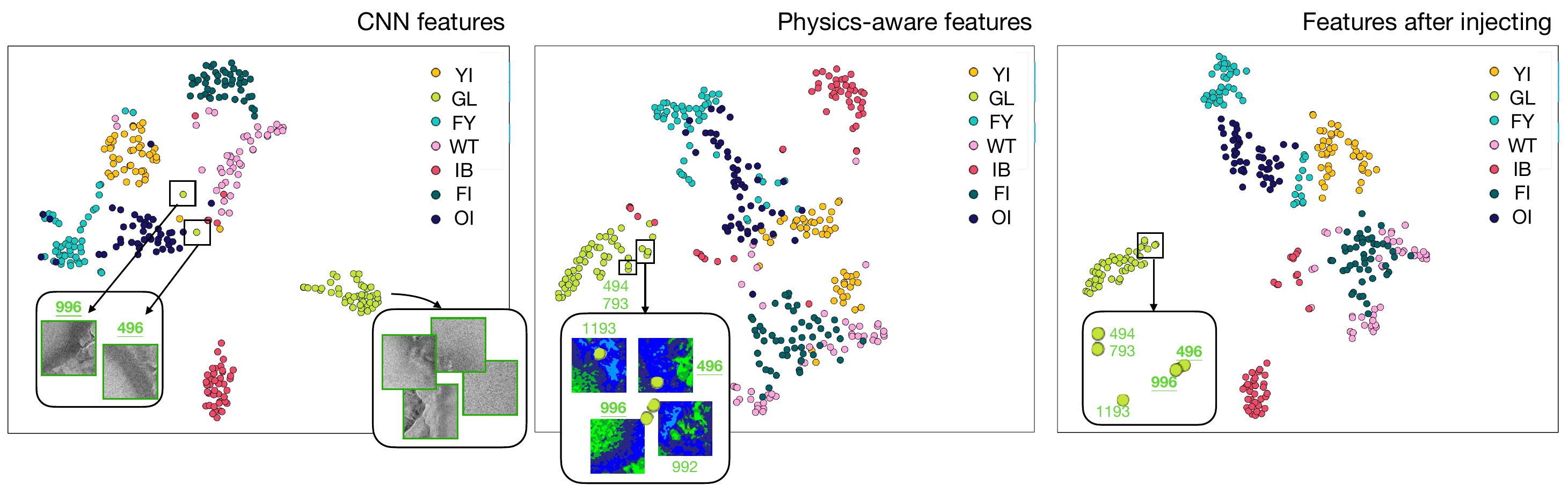}
         \caption{}
     \end{subfigure}
     ~
     \\
     \begin{subfigure}[b]{0.95\textwidth}
         \centering
         \includegraphics[width=0.9\textwidth]{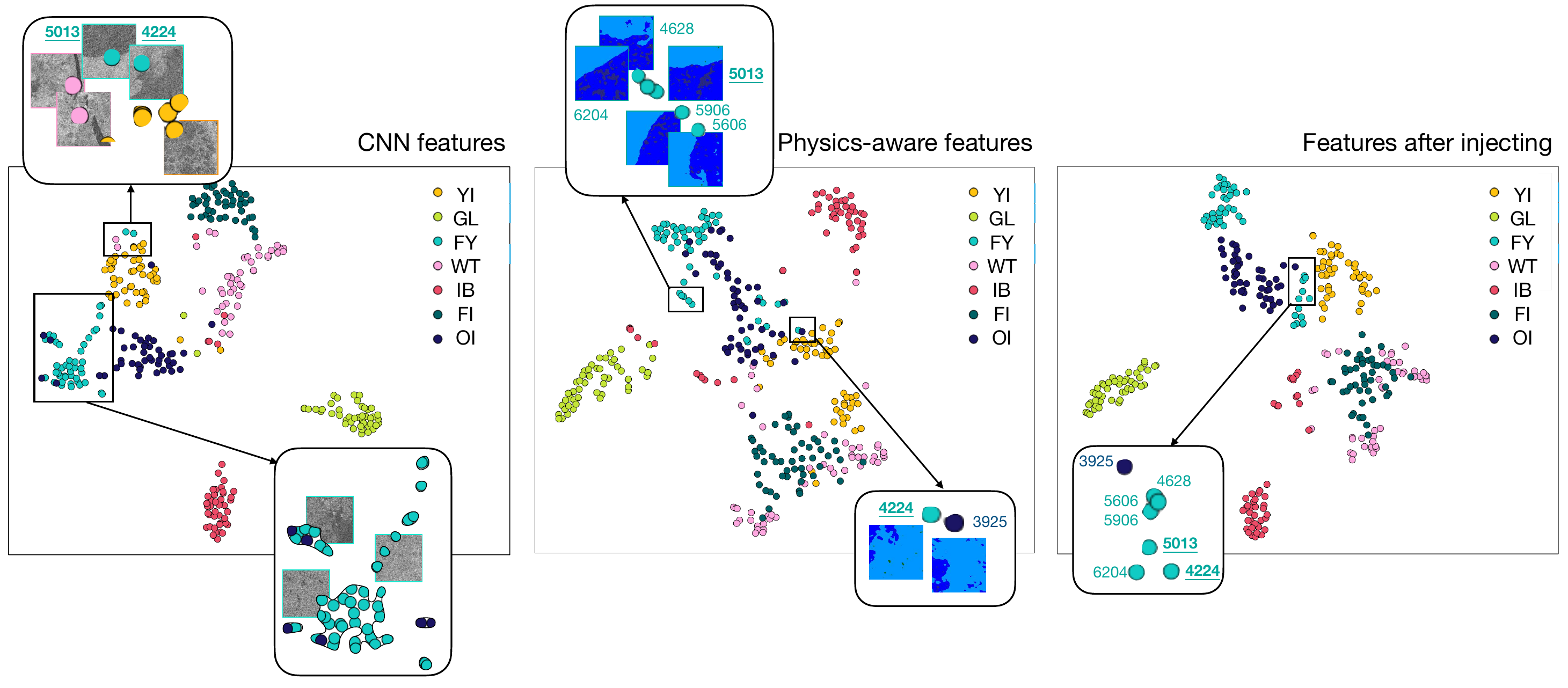}
         \caption{}
     \end{subfigure} 
     \caption{The feature visualization of CNN supervised training result, the physics guided learning result, and the result after physics injected learning.}
     \label{fig_explain}
\end{figure*}

Two cases are presented in Fig. \ref{fig_explain} to demonstrate how the PIN preserves the physics consistency of features. We visualize the features of CNN supervised training, physics-aware features from unsupervised PGN, and the features after injection in PIN by t-sne \cite{maaten2008visualizing}. As shown in Fig. \ref{fig_explain}(a), the CNN feature of instance 996 and 496 in glacier class are far away from the majority due to the different image contents. Based on the similar physics attributes, the middle figure in Fig. \ref{fig_explain}(a) shows the two samples are close to most glacier data in physics-aware features. In the visualization of features after physics injected learning, we can observe sample 496 and 996 are still close to each other due to the similar visual representation, and also, they maintain the closeness with sample 494, 793, and 1193 as they were in physics-aware features. As a result, we infer the physics injected learning can preserve the physics consistency during the network training. Another example is about sample 5013 and 4224, shown in Fig. \ref{fig_explain}(b). They have similar image features in CNN training while different from the other first year ice samples. The physics-aware features in the middle figure reveals that sample 5013 and 4224 have their own physics characteristics. The physics constraints existed in physics-aware features are, i.e, sample 5013 having similar physics properties with 4628, 5606, 5906, and sample 4224 being very close to sample 3925 from old ice class. PIN continues this kind of constraint, as shown in the right figure of Fig. \ref{fig_explain}(b). In a word, the proposed PGN and PIN represent the SAR images from a more comprehensive perspective than the traditional supervised CNN learning.

\subsubsection{The Inspiration from Explainability}

In addition, the explainability we discussed before can inspire us to improve our deep learning algorithm in the future work. For example, the above analysis indicates the floating ice has a similar physics BoT representation with water bodies so that the physics-aware features of them are not well discriminative in semantic level, as shown in Fig. \ref{fig_explain}. On the contrary, they can be discriminative based on the visual contents. The final PIN result shows injecting physics-aware features could not improve the performance of recognizing water bodies and floating ice, because their physics knowledge is not as helpful as other classes. We can see in Fig. \ref{fig_cmatrix} that the true positive samples of water bodies and floating ice in PGIL result are fewer than those in CNN training result. Thus, it inspires us to re-think the physics injected learning strategy that the constraint of physics consistency should be relaxed in such classes. In our future work, we will further improve the physics injected learning method in this direction to achieve a better result.

\section{Conclusion}
\label{sec:con}

In this paper, we propose a novel physics guided and injected learning neural network for SAR image classification with limited labeled data, to explore the potential of physically explainable deep learning. Three components of PGIL are explainable model, physics guided network, and physics injected network. The prior knowledge in explainable models is encoded into the physics-aware features via unsupervised PGN learning, and then is injected in the classification pipeline through PIN, supervised by limited labeled data. The hybrid Img-Phy dataset format is proposed for evaluation and abundant experiments are conducted on Sentinel-1 and Gaofen-3 SAR data. The results demonstrate the semantic discrimination and the physics awareness of the learned features by PGN, as well as the good generalization. Additionally, we discuss the interpretability of the guided signals in the established surrogate task to prove the results are with physical constraint. The advantages of the proposed PGIL are (1) the unsupervised PGN is a plug-and-play module which can be integrated to any deep learning framework for physics-aware feature injection; (2) the physics knowledge injection is capable of preserving the physics consistency in the prediction and preventing overfitting in case of limited labeled data; (3) the results are explainable with the help of the distinct physical models and expertise to a certain extent, that inspire us to further improve the deep learning model in the right direction. The sea-ice dataset and source code are publicly in \url{https://github.com/Alien9427/XAI4SAR-PGIL}.

\section{Acknowledgment}

This work was supported in part by the National Natural Science Foundation of China under Grant 62101459, U20B2068, and in part by the China Postdoctoral Science Foundation under Grant BX2021248, and the Fundamental Research Funds for the Central Universities under Grant G2021KY05104.

\bibliographystyle{cas-model2-names}

\bibliography{My12paper_mybibfile}



\end{document}